\makeatletter
\declare@file@substitution{revtex4-1.cls}{revtex4-2.cls}
\makeatother

\documentclass[preprint]{aastex631}

\usepackage{natbib}
\usepackage{url}
\usepackage{xcolor}  
\usepackage{lineno}

\begin{document}

\title{The Double-lined Eclipsing $\gamma$ Doradus System AX Draconis in a 0.568-day Orbit }
\correspondingauthor{Jae Woo Lee}
\email{jwlee@kasi.re.kr}
\author[0000-0002-5739-9804]{Jae Woo Lee}
\affil{Korea Astronomy and Space Science Institute, Daejeon 34055, Republic of Korea}

\author[0000-0002-8394-7237]{Min-Ji Jeong}
\affil{Korea Astronomy and Space Science Institute, Daejeon 34055, Republic of Korea}

\author[0000-0002-8692-2588]{Kyeongsoo Hong}
\affil{Korea Astronomy and Space Science Institute, Daejeon 34055, Republic of Korea}

\author[0000-0001-9339-4456]{Jang-Ho Park}
\affil{Korea Astronomy and Space Science Institute, Daejeon 34055, Republic of Korea}

\author[0000-0003-1916-9976]{Pakakaew Rittipruk}
\affil{National Astronomical Research Institute of Thailand, Chiang Mai 50200, Thailand}

\author[0000-0002-6687-6318]{Hye-Young Kim}
\affil{Korea Astronomy and Space Science Institute, Daejeon 34055, Republic of Korea}

\begin{abstract} 
For the near-contact binary AX Dra, we present the first time-series spectroscopy collected with the echelle spectrograph BOES. 
From spectral analysis, we measured the projected rotation of $v_1 \sin i$ = $120\pm21$ km s$^{-1}$ and effective temperature 
of $T_{\rm eff,1}$ = $7220\pm150$ K for the brighter primary component, together with radial velocities (RVs) for both stars. 
To obtain a consistent binary model, the RV curves were analyzed by combining the 2-min cadence photometric data observed in 
the TESS sectors 15, 21, 22, and 41. The modeling indicates that AX Dra is a semi-detached system exhibiting 
a total secondary eclipse, with the detached primary component having a large filling factor of 92 \%. The system has masses 
of $1.717\pm0.026$ $M_\odot$ and $0.804\pm0.014$ $M_\odot$, radii of $1.541\pm0.020$ $R_\odot$ and $1.237\pm0.014$ $R_\odot$, 
luminosities of $5.78\pm0.50$ $L_\odot$ and $0.83\pm0.05$ $L_\odot$, and a temperature difference of 
$\Delta$($T_{\rm eff,1}$--$T_{\rm eff,2}$) = $2263\pm163$ K. Multi-frequency analyses of the TESS residual lights yielded 
35 significant signals in the frequency range below 5 day$^{-1}$. Among them, four frequencies of $f_1$, $f_2$, $f_3$, and $f_5$ 
are independent $\gamma$ Dor pulsations of the primary star, for which two acceptable mode-identification solutions were obtained 
using the frequency ratio method. These results suggest that AX Dra is the shortest-period double-lined eclipsing binary containing 
a $\gamma$ Dor-type pulsator and that the pulsating primary is likely an accretor affected by mass transfer. 
\end{abstract}


\section{INTRODUCTION} \label{sec_intro}

Asteroseismology is used to probe the interior physics of pulsating stars, from the near-core region to the outer envelope, 
ultimately to better understand stellar evolution. $\delta$ Sct and $\gamma$ Dor stars are intermediate-mass stars with 
masses of 1.2$-$2.5 $M_\odot$ and spectral types of A$-$F, that display multi-period pulsations \citep{Antoci+2019}. 
Their positions on the Hertzsprung--Russell (H--R) diagram are at the intersection of the Cepheid instability strip and 
the main sequence band, and the blue edge of the $\gamma$ Dor domain overlaps with the red edge of the $\delta$ Sct one. 
Space-based ultra-precise photometric surveys such as the {\it Kepler} mission indicate that many pulsating variables are 
hybrid pulsators with two detected frequency regions, one lower ($\gamma$ Dor) and one higher ($\delta$ Sct) than 
$\sim$5 day$^{-1}$ \citep[e.g.][]{Grigahcene+2010}. These two types of pulsators exhibit significant differences in 
their observed oscillation periods ($P_{\rm pul}$) and pulsation constants ($Q$) \citep{Breger2000,Handler+Shobbrook2002}. 

Double-lined eclipsing binaries (DEBs) are nearly the only stars whose fundamental parameters, such as masses and radii, can 
be directly measured from radial velocities (RVs) and light curves. The absolute dimensions can be measured with a precision 
of better than 3 \% and are the basis for understanding stellar physics \citep{Torres+2010}. DEBs that show multiple pulsations 
are of great value, because the synergy from both eclipses and pulsations can help improve the models of stellar structure and 
evolution \citep{Kurtz2022}. We have been performing medium- and high-dispersion spectroscopic observations of pulsating EBs 
using 2-m class telescopes \citep{Hong+2015,Rittipruk+2025}. 

In this article, we focus on the EB system AX Dra (TIC 148963419; TYC 4168-362-1; Gaia DR3 1680579406815027584; 
$V$ = $+$10.90, $(B-V)$ = $+$0.40), which was reported to be a possible variable by \citet{Kippenhahn1955} and proposed 
by \citet{Strohmeier+Knigge1961} as a binary with an eclipsing period of about 0.568 days. \citet{Kim+2004} reviewed 
the historical details of this system prior to 2004, and conducted photometric studies from their multiband observations 
during the 2001 and 2003 seasons and all available minimum times. As a consequence, they classified AX Dra as 
a near-contact EB with a mass ratio of $q$ = 0.629, an inclination angle of $i$ = 82.8 deg, effective temperatures of 
$T_{\rm eff,1}$ = 6850 K and $T_{\rm eff,2}$ = 4950 K, and a light contribution of $l_1/(l_1+l_2)$ = 0.85 in the $V$ band, 
and interpreted the seasonal light changes as a variable cool spot on the secondary companion. In addition, \citet{Kim+2004} 
reported a cyclical variation with a period of 56 years and a semi-amplitude of 0.006 days from the eclipse timing $O-C$ diagram, 
and suggested that the light-travel-time effect caused by a tertiary companion ($M_{3} \sin i_{3} = 0.13 M_\odot$) could 
explain the observed variation. Since then, the short-period EB has been a quite neglected stellar system. 

AX Dra is one of the spectral type A$-$F stars observed with the TESS's 2-min cadence. These observations imply that 
the TESS target could be a pulsating EB candidate. Here, we present the first spectroscopic observations of AX Dra, and 
the results of its binary modeling and pulsation frequency analysis.

\section{TESS PHOTOMETRY, NEW ECLIPSE TIMES, AND LIGHT VARIATION} \label{sec_photometry}   

AX Dra was observed as part of the TESS mission to find small exoplanets using the transit method. The 2-min sampling data of 
the program target were obtained during four non-consecutive sectors 15, 21, 22, and 41 between August 2019 and August 2021. 
The CROWDSAP factor over these sectors is 0.9924$\pm$0.0013 on average \citep{Lee+2022}. In this study, we took the stellar 
fluxes from simple aperture photometry (\texttt{SAP$_-$FLUX}) available at the MAST\footnote{\url{https://archive.stsci.edu/}. 
The observations can be found at \url{http://dx.doi.org/10.17909/1536-ze66}.}. 
The detrending and normalization of the TESS data are identical to those in \citet{Lee+2019}. The corrected fluxes were 
transformed to magnitudes by requiring an average brightness at maximum light to be $T\rm_P$ = $+$10.62 \citep{Paegert+2022}. 
In all, 74,036 observations were used for our analysis, with a time span of about 735 days (BJD 2,458,711.36 $-$ 2,459,446.58). 
They are shown as magnitude versus BJD on the top panel in Figure~\ref{Fig1}, where the individual observations are represented 
by different colored circles for each sector. 

A total of 362 mid-eclipse times and their associated errors (181 for each eclipse type) were measured using the TESS observations 
\citep{Kwee+Van1956}. These appear in Table~\ref{Tab1}, in which the timings of the primary and secondary eclipses are indicated 
by Min I and Min II, respectively. We used all primary minima in a linear least-squares solution to get a new light ephemeris 
(epoch $T_0$ and period $P_{\rm orb}$) suitable for the TESS photometric and our spectroscopic data, as follows:
\begin{equation}
\mbox{Min I} = \mbox{BJD}~ 2,458,711.514732(\pm0.000029) + 0.568163473(\pm0.000000045)E.  
\end{equation}
The error of each coefficient is one standard deviation (1$\sigma$). The phase-folded light curve of AX Dra is illustrated on 
the second panel in Figure~\ref{Fig1}. Our new $P_{\rm orb}$ is equivalent to the binary star's frequency of $f_{\rm orb}$ = 1.7600568 $\pm$ 0.0000001 day$^{-1}$. 

To find out the light variations of AX Dra, we measured the light levels in magnitudes at two light maxima (Max I and Max II) and 
two eclipse minima (Min I and Min II) for each TESS sector. A summary of these measurements is presented in Table~\ref{Tab2}. Here, 
the BJD time is the median between the first and last observations used in each dataset. The magnitude differences between 
the light levels were computed and they were plotted in Figure~\ref{Fig2}. As indicated in this figure, AX Dra exhibits brightness 
variations over time, which are most likely caused by temporal changes in starspot activity.

\section{GROUND-BASED SPECTROSCOPY AND SPECTRAL ANALYSIS} \label{sec_spectroscopy}

To help determine the reliable fundamental parameters for AX Dra, we conducted phase-resolved spectroscopy using the echelle 
spectrograph BOES, which is fiber-fed to the BOAO 1.8-m telescope \citep{Kim+2007}. A total of 46 spectra were collected over 
eight nights in 2022 June and November. The spectral resolution was selected to be $R$ = 30,000 and the data were composed of 
75 orders in the wavelength range of 3600$-$10,200 $\rm \AA$. Each exposure time was set to 960 s, no more than 0.02 of 
the eclipsing period. This indicates that orbital smearing could be neglected. The signal-to-noise (S/N) ratios for our target star 
were typically between 30 and 50 on a continuum around the \ion{Mg}{1b} triplet region. All observed spectra were pre-processed 
(bias, flat-field corrections) and calibrated (wavelength calibration, normalization) using the \texttt{IRAF/ECHELLE} task.

\citet{Kim+2004} reported that the short-period eclipsing system AX Dra has a large difference of about 1900 K in temperature 
between the components, and the faint secondary makes only a small contribution of about 15 \% to the overall binary light in 
the $V$ band. For the components' RV measurements, we constructed the phase-folded trailed spectra over the entire wavelength and 
looked for the S-wave features resulting from the binary orbital motion \citep[see][]{Lee+2018}. As a consequence, 
measurable absorption lines were found in the \ion{Mg}{1b} triplet region ($\lambda$5167.33, $\lambda$5172.70, and $\lambda$5183.62). 
The \ion{Mg}{1b} triplets were broadened by the fast rotation of each star and blended with each other by the orbital motion. 
Thus, the RVs for both components were derived using the broadening function \citep{Rucinski1992,Rucinski1999,Rucinski2002} in 
the \texttt{RAVESPAN} software \citep{Pilecki+2017}. In the runs, the template spectra were imported from the model library made 
available by \citet{Coelho+2005} considering their temperatures. The resulting double-lined RVs are presented in Table~\ref{Tab3} 
and Figure~\ref{Fig3}.

For the projected rotation ($v\sin$$i$) and surface temperature ($T_{\rm eff}$) determination, the disentangling spectra of 
AX Dra AB were constructed employing the \texttt{FDBinary} program\footnote{\url{http://sail.zpf.fer.hr/fdbinary}} 
\citep{Ilijic+2004}. However, the reconstructed spectrum of the cool companion was insufficient to estimate the reliable values 
of the two atmospheric parameters, because its light contribution and the S/N ratio were both low. The reconstructed spectrum of 
the hotter primary was analyzed by means of the \texttt{iSpec} code \citep{Blanco-Cuaresma+2014,Blanco-cuaresma2019} 
\footnote{\url{https://www.blancocuaresma.com/s/iSpec}}, which calculates the atmospheric parameters by minimizing 
the $\chi^2$ difference between observations and synthetic models. For the synthetic spectra, we adopted the \texttt{SPECTRUM} 
radiative transfer code \citep{Gray+Corbally1994}, the ATLAS9 stellar model atmospheres \citep{kurucz2005}, and the atomic line list 
version 5.0\footnote{\url{http://ges.roe.ac.uk/docs/GES\_linelist\_v5\_report.pdf}} from the Gaia-ESO survey \citep{Heiter+2015}. 
Also, the surface gravity of $\log g_1$ = 4.30 (see Table~\ref{Tab5}) and the solar abundance were used for the model spectra. 

The nine spectral regions of \ion{Ca}{1} $\lambda$4226, \ion{Fe}{1} $\lambda$4250, \ion{Fe}{1} $\lambda$4271, \ion{Cr}{1} $\lambda$4290, 
H$_{\rm \gamma}$, \ion{Y}{2} $\lambda$4376, \ion{Fe}{1} $\lambda$4383, H$_{\rm \beta}$, and \ion{Mg}{1b} triplet were chosen 
from the Stellar Spectral Classification Atlas by \citet{Gray+Corbally2009}. 
These absorption lines help classify the star temperatures of spectral types A$-$F. Finding the optimal atmospheric parameters 
resulted in $v_1\sin$$i$ = $120\pm21$ km s$^{-1}$ and $T_{\rm eff,1}$ = $7220\pm150$ K. The micro- and macro-turbulent velocities 
were calculated to be $v_{\rm{mic}}$ = 1.88 km s$^{-1}$ and $v_{\rm{mac}}$ = 17.6 km s$^{-1}$ from the \texttt{iSpec} code, respectively. 
The chosen spectral regions of the disentangling spectrum are plotted with the best-fit model spectrum in Figure~\ref{Fig4}.

\section{BINARY MODELING AND ABSOLUTE DIMENSIONS} \label{sec_binarity}

As presented in Section 2, the TESS observations of AX Dra indicate that the target's light has continuously varied, as was 
especially prominent during primary eclipse. During this minimum phase, the pulsating primary star is deeply eclipsed, and 
the companion star dominates the observed light. Given the secondary’s temperature and the starspot activity suggested by 
previous studies (\citep{Kim+2004}), the enhanced variation during primary eclipse is likely associated with time-dependent 
brightness modulation caused by spot activity on the secondary component. In addition, the light curve shows a depth difference 
of about 0.65 mag between the two eclipses, while the secondary minimum appears flat-bottomed. These features indicate that 
the more massive primary component (AX Dra A) being eclipsed at Min I is hotter and larger than its companion (AX Dra B). 
To derive a unique solution for AX Dra, our spectroscopic measurements were simultaneously solved with all archival TESS data 
observed at 2-min cadence. For the combined modeling of these data, we introduced the \texttt{Wilson-Devinney} (W-D) code 
\citep{Wilson+Devinney1971,Van+Wilson2007} that adequately models the geometrical distortion of each star and permits 
proximity effects for RVs. Here, the subscripts 1 and 2 refer to AX Dra A and B, respectively. 

The surface temperature of the AX Dra primary was set to the value of $T_{\rm eff,1}$ = 7220$\pm$150 K, based on 
our spectral analysis (Section~\ref{sec_spectroscopy}). Both bolometric albedo ($A_1$) and gravity-darkening ($g_1$) parameters 
for AX Dra A were held to typical values of 1.0, while the $A_2$ and $g_2$ values for AX Dra B were fitted during our W-D run. 
The logarithmic law was applied for the non-linear limb darkening, and the $x$ and $y$ coefficients were initialized with 
updated values of \citet{Van1993}. With the measured rotation of $v_1$$\sin$$i$ = $120\pm21$ km s$^{-1}$, 
the rotation parameter of AX Dra A was assigned as $F_1$ = 0.87$\pm$0.15. For the lobe-filling companion, a synchronous rotation 
of $F_2$ = 1.0 was used. We performed binary star modeling using the semi-detached and contact modes of the W-D program. 
The results confirmed that AX Dra is in a semi-detached configuration, with the lower-mass component filling its Roche lobe. 
The binary model parameters are summarized in Table~\ref{Tab4}, where each parameter's error was obtained by applying 
the error estimation method of \citet{Southworth+Bowman2022}. The light and RV curves from our solutions are depicted as 
red solid lines in Figures~\ref{Fig1} and \ref{Fig3}. 

Detailed analyses combining our echelle spectra with space-based photometric data indicate that AX Dra is a short-period EB 
with a temperature difference of 2263 K and a primary-star fill-out factor of 92 \%. Previous studies have shown that 
short-period EBs with intermediate-mass pulsators can exhibit a broad range of Roche-lobe filling factors \citep[e.g.][]{Zhang+2013}, 
while temperature differences between components may also vary substantially depending on the binary configuration. In this context, 
AX Dra is particularly noteworthy as a strongly interacting near-contact binary that provides an interesting environment for 
investigating the pulsational properties discussed in the following sections. 

The combined model parameters listed in Table~\ref{Tab4} are well suited to provide direct and accurate absolute dimensions of 
the binary star. This calculation process is the same as that previously performed by \citet{Lee+2018,Lee+2021}. The resultant 
parameters are presented in Table~\ref{Tab5}. To compute the distance to AX Dra, we used the apparent magnitude and color index 
of the system to be $V$ = +10.90$\pm$0.06 and ($B-V$) = +0.40$\pm$0.08 \citep{Hog+2000}, respectively. The intrinsic color index 
was estimated to be ($B-V$)$\rm_0$ = +0.30$\pm$0.03 from $T_{\rm eff,1}$ = 7220$\pm$150 K \citep{Flower1996,Pecaut+Mamajek2013}, 
and then derived $E$($B-V$) = $+$0.10. Finally, our obtained photometric-spectroscopic distance was 381$\pm$19 pc, which agrees 
well with 388$\pm$2 pc measured from the Gaia trigonometric parallax of 2.580$\pm$0.011 mas \citep{Gaia2023}.

\section{PULSATION FREQUENCY ANALYSIS} \label{sec_pulsation}

\begin{figure}
\includegraphics{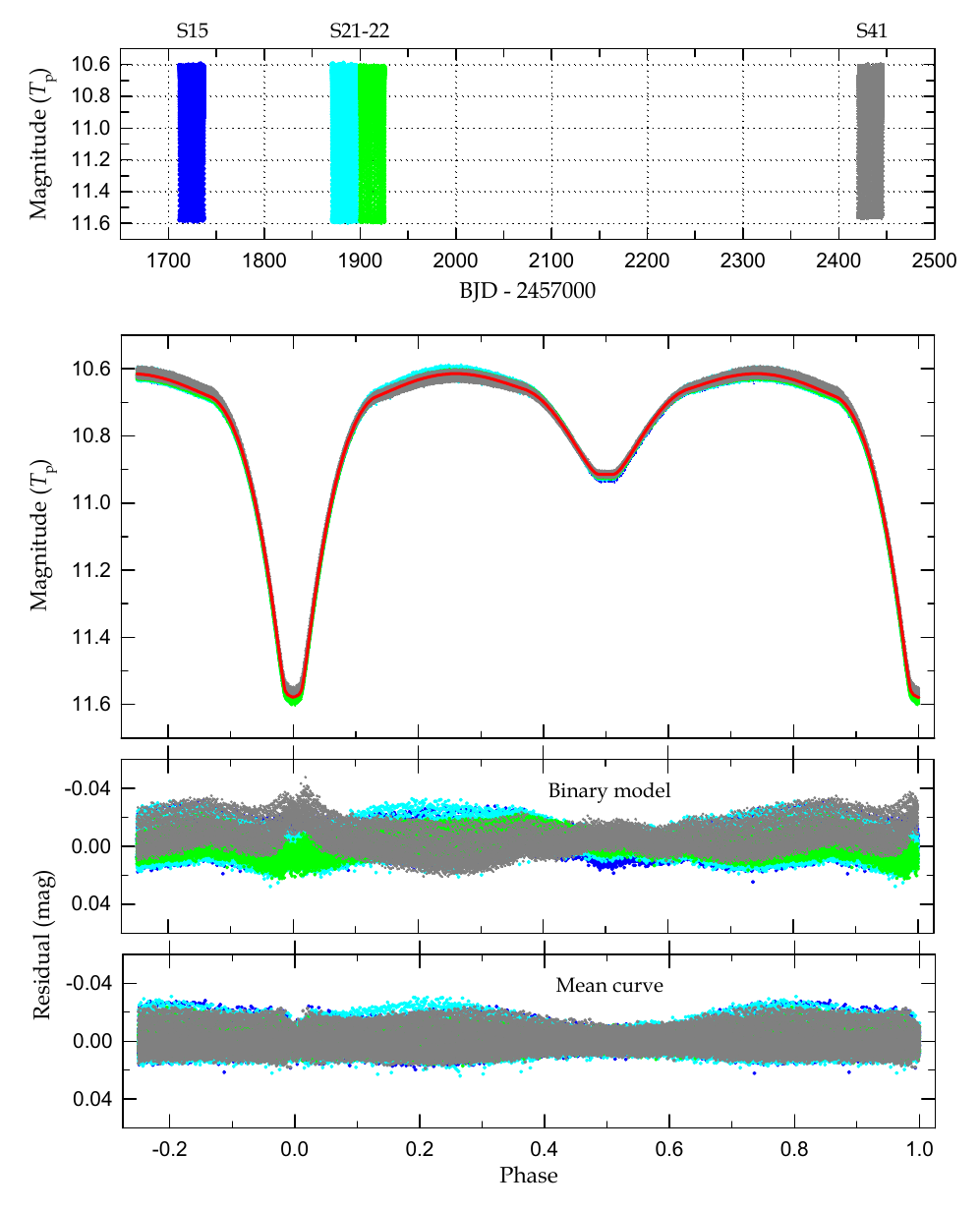}
\caption{TESS observations of AX Dra distributed in BJD (top panel) and orbital phase (second panel). Different colored circles 
for each sector are individual measures, and the red solid curve represents the model curve obtained with our W-D synthesis. 
The third and bottom panels show the corresponding residuals from the binary model and from each sector's mean curve with a phase 
interval of 0.002, respectively. }
\label{Fig1}
\end{figure}

\begin{figure}
\includegraphics[scale=1.0]{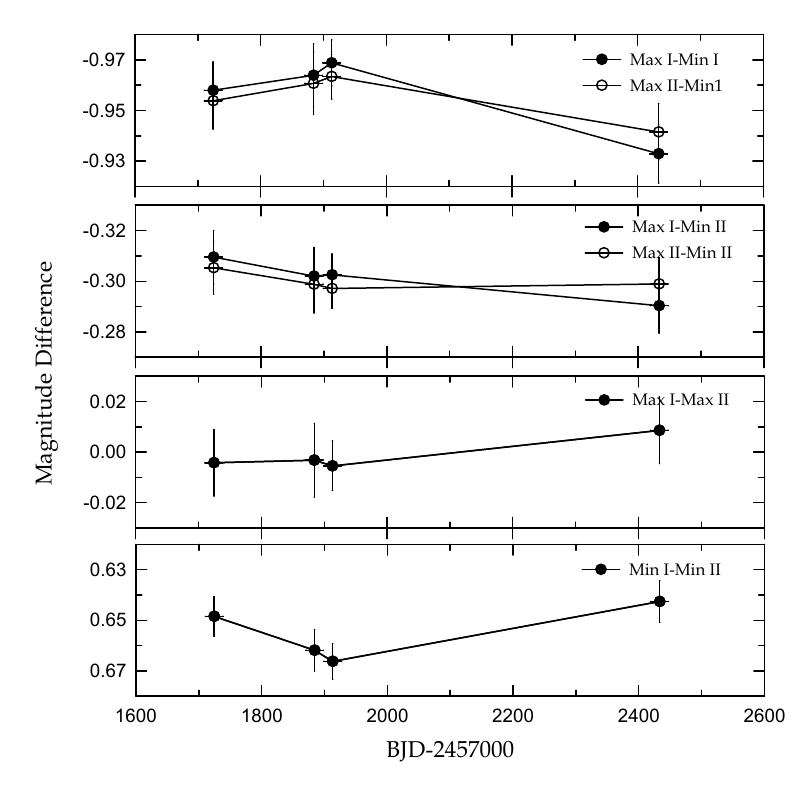}
\caption{Variations in magnitude differences with time between four characteristic phases (Max I, Min I, Max II and Min II) for AX Dra. }
\label{Fig2}
\end{figure}

\begin{figure}
\includegraphics{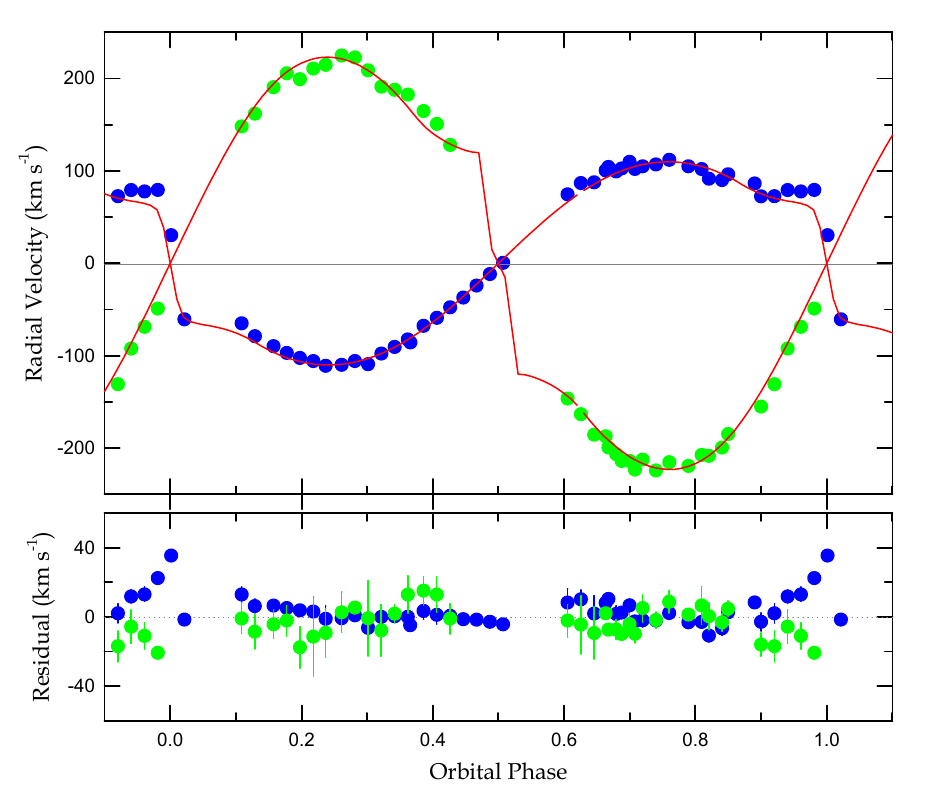}
\caption{RV curves of AX Dra with fitted models. The blue and green circles are our measurements for the primary and secondary 
stars, respectively, and the red solid curves represent the results from a consistent light and RV curve analysis with the W-D code. 
The gray line in the upper panel denotes the system velocity of $-$1.5 km s$^{-1}$. The lower panel shows the residuals between 
observations and models. }
\label{Fig3}
\end{figure}

\begin{figure}
\includegraphics{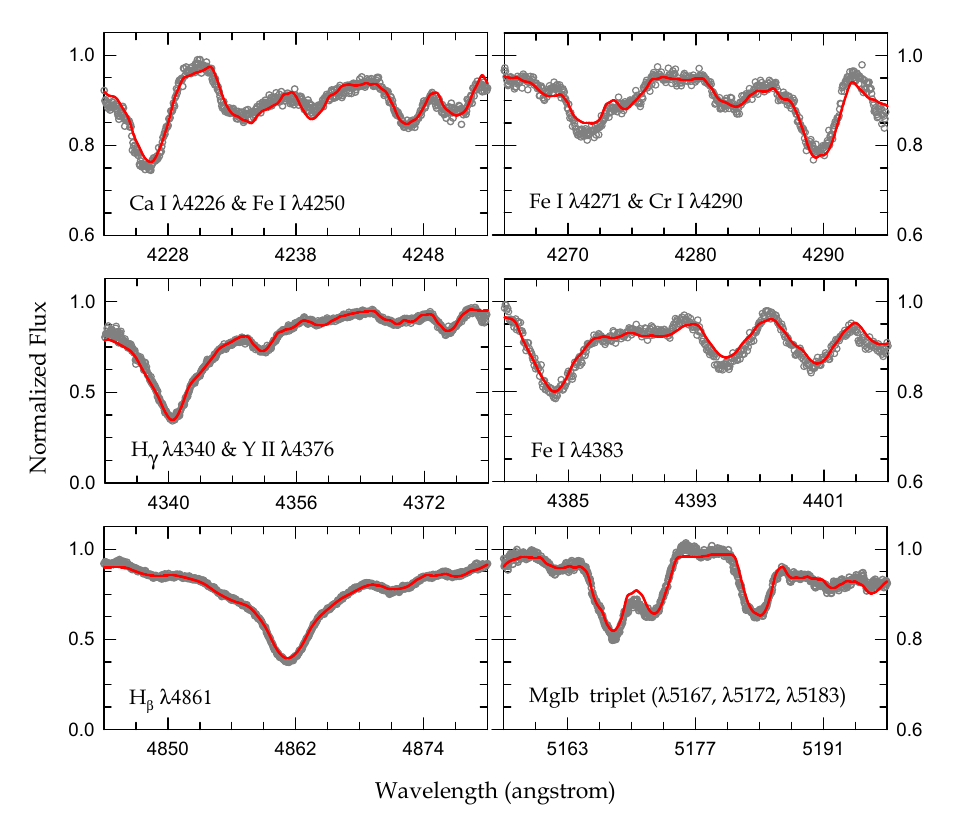}
\caption{Selected spectral regions of the primary star AX Dra A. The gray circles are the disentangling spectrum obtained with 
the \texttt{FDBinary} code. The red lines denote the synthetic spectrum with $T_{\rm eff,1}$ = 7220 K, $\log$ $g_1$ = 4.30, 
and, $v_1$$\sin$$i$ = 120 km s$^{-1}$. } 
\label{Fig4}
\end{figure}

The fundamental parameters in Table~\ref{Tab5} suggest that the AX Dra A primary is an F0-type main-sequence star located where 
the $\delta$ Sct and $\gamma$ Dor instability regions on the H--R diagram overlap each other \citep[e.g.][]{Uytterhoeven+2011,Xiong+2016}. 
Then, this object would be a candidate for a hybrid pulsator, which is expected to show both pressure ($p$) and gravity ($g$) modes. 
In the third panel of Figure~\ref{Fig1}, the residual lights from our W-D model exhibit a rather large scatter of $\pm$0.02 mag. 
This scatter is primarily attributed to multiple oscillations with a characteristic period of roughly 0.47 days. In addition to 
the eclipse and pulsation effects, the TESS observations may be convoluted with light changes due to local surface inhomogeneities, 
such as starspots. Because of these, there is a notable discrepancy in brightness between the four sectors as shown in Figure~\ref{Fig1}. 

\begin{figure}
\includegraphics[]{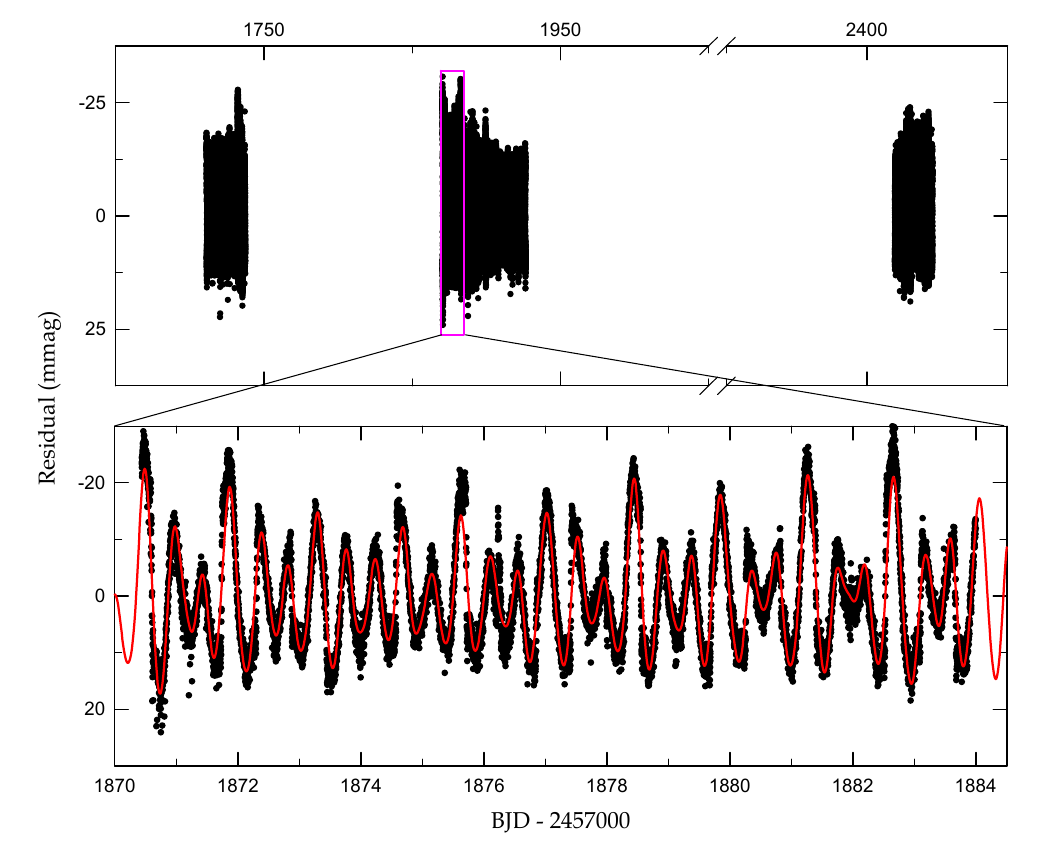}
\caption{Light curve residuals from the mean curves distributed in BJD. The lower panel presents a short section of the residuals 
marked using the inset box of the upper panel. The red synthetic curve is computed from the 35-frequency fit to the data between 
orbital phases 0.13 and 0.87. }
\label{Fig5}
\end{figure}

For better analysis of the pulsations, we made the normal points for each sector from the observed TESS data, evenly binned with 
a phase interval of 0.002 \citep{Lee+2023}. The individual residuals after removing the mean curves were used to obtain amplitude 
spectra up to the Nyquist limit of 360 day$^{-1}$ with the \texttt{PERIOD04} \citep{Lenz+Breger2005}. The mean-curve residuals 
are plotted against phase in the bottom panel of Figure~\ref{Fig1} and BJD in Figure~\ref{Fig5}. 

\begin{figure}
\includegraphics[]{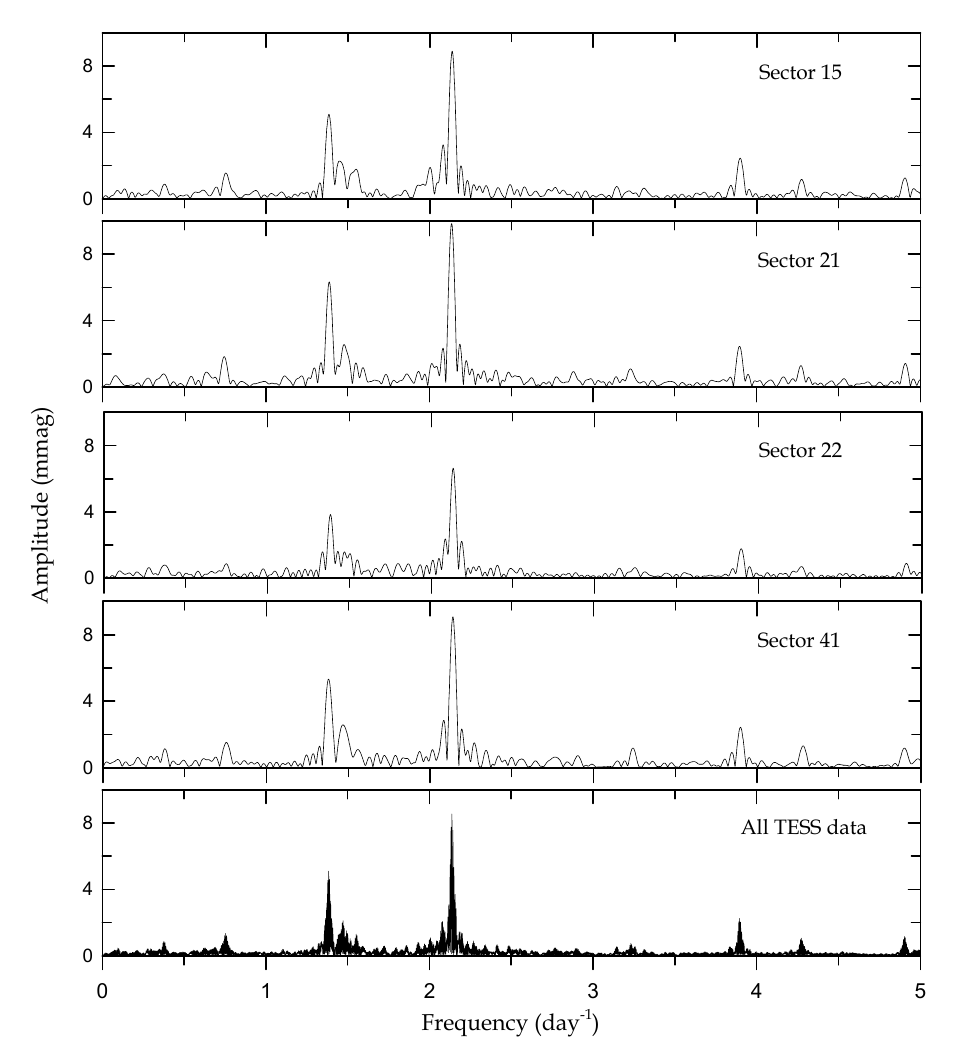}
\caption{\texttt{PERIOD04} periodograms for the light curve residuals except for the primary eclipse phase. The first to fourth panels 
show the amplitude spectra for the residuals in each sector, while the bottom panel is obtained using the entire TESS data. }
\label{Fig6}
\end{figure}

First of all, following the pre-whitening process conducted by \citet{Lee+2014}, we separately analyzed each sector's residuals 
in the orbital phases 0.13$-$0.87, excluding the data from the primary eclipses. The results are illustrated in the first to 
fourth panels of Figure~\ref{Fig6} and reveal the presence of multi-period pulsations having dominant signals at $\sim$1.38 day$^{-1}$ 
and $\sim$2.14 day$^{-1}$. No conspicuous signals were seen in the frequency domain greater than 5 day$^{-1}$. These imply 
that AX Dra A is a $\gamma$ Dor-like pulsator. Then, we applied the \texttt{PERIOD04} software to the secondary- and outside-eclipse 
residuals of the entire TESS data, and detected a total of 35 frequency signals \citep{Breger+1993} presented in Table~\ref{Tab6}. 
Of the extracted frequencies, possible combinations were found within 1.5 times the Rayleigh frequency of 1/$\Delta T$ 
\citep{Loumos+Deeming1978}. Here we used a time span of $\Delta T$ = 215 days, excluding S41 with a long observation gap of more than 
one year since S22 \citep{Breger+Bischof2002}. The search result is presented in the remark column of Table~\ref{Tab6}. Seven frequencies, 
denoted by superscript $c$ in this table, were detected in all four sectors. We think the four significant signals ($f_1$, $f_2$, $f_3$, $f_5$) 
except for combination frequencies are considered independent oscillating modes excited in AX Dra A.

\section{DISCUSSION AND CONCLUSIONS} \label{sec_conclusion}

In 2021, we identified the near-contact system AX Dra as a possible pulsating EB using high-cadence (2 min) TESS data. 
To characterize its physical properties in detail, we subsequently included this target in our observing program and performed 
high-dispersion spectroscopic observations with BOES on the 1.8-m telescope. From the echelle spectra, 
the double-lined RVs were measured and the atmosphere parameters of AX Dra A were determined to be $v_1\sin$$i$ = 120$\pm$21 km s$^{-1}$ 
and $T_{\rm eff,1}$ = 7220$\pm$150 K. Our spectroscopic measurements were modeled with the ultra-precise photometric data obtained at 
four TESS sectors. The synthetic modeling directly calculated the fundament parameters of the binary target without making any assumptions. 
The masses, radii, and temperatures for each component are accurate to within 2 \%. The EB-based distance of 381$\pm$19 pc yields 
the parallax of 2.62$\pm$0.13 mas, which is well-matched with the Gaia value of 2.580$\pm$0.011 mas \citep{Gaia2023}. 

The projected rotational velocity, $v_1 \sin i$, is effectively identical to the true equatorial velocity ($v_1$) given the nearly 
edge-on configuration of the system ($i$ = 89.07 deg). The measured value is marginally lower than the synchronous one of $v_{\rm 1,sync}$ 
= 137.3$\pm$1.8 km s$^{-1}$, suggesting subsynchronous rotation. This departure from synchronous rotation is uncommon in short-period 
semi-detached binaries like AX Dra, where strong tidal interactions are expected to enforce rapid spin–orbit synchronization. 
While angular momentum loss via magnetic braking and differential rotation due to structural changes may contribute to this behavior 
\citep{Lurie+2017,Fleming+2019,Koenigsberger+2021,Hobson-Ritz+2025}, other mechanisms should also be considered. 
In particular, ongoing or recent mass transfer and the associated redistribution or loss of angular momentum within the system may 
lead to temporary departures from synchrony \citep{Sepinsky+2007,Sepinsky+2010,Tauris+van2006}. Alternatively, the presence of 
a tertiary companion and its dynamical interaction with the inner binary could also induce subsynchronous rotation \citep{Fuller+Felce2023}. 
Indeed, \citet{Kim+2004} reported a quasi-sinusoidal variation in the eclipse timing residuals and suggested 
a possible light-travel-time effect caused by a tertiary body. Although our spectroscopic analysis and combined light-curve modeling reveal 
no significant evidence for a third component, the existence of a faint low-mass circumbinary companion cannot be completely excluded. 

Time-series photometry from TESS reveals temporal variations in the brightness of AX Dra. These luminosity changes are likely produced 
by a combination of pulsations and additional effects such as starspots. To extract oscillating frequencies more reliably, we generated 
the mean light curves for each sector using a phase bin of 0.002 and subtracted them from the observed TESS data. The resulting residual 
lights were then analyzed in detail. We extracted 35 significant signals, among which the four frequencies of $f_1$, $f_2$, $f_3$, and 
$f_5$ were classified as possible pulsation modes originating from the AX Dra primary, located in the instability regions of A$-$F dwarfs 
on the H--R diagram. Using the frequency (period)-density relation of $Q_i$ = $f_i ^{-1}$$\sqrt{\rho_1 / \rho_\odot}$, we computed the pulsation 
constants $Q$ for the low frequencies presented in the third column of Table~\ref{Tab7}. The oscillating signals of 1.383$-$2.137 day$^{-1}$ 
and the $Q$ values of 0.320$-$0.494 days are typical of $\gamma$ Dor-type $g$ modes with frequencies of $f <$ 5 day$^{-1}$ and pulsation 
constants of $Q >$ 0.23 days \citep{Henry+2005,Uytterhoeven+2011}. 

The radial order ($n$) and spherical degree ($\ell$) of the $g$-mode frequencies can be identified by means of the frequency ratio 
method (FRM) presented in \citet{Moya+2005} and applied in our previous studies \citep{Lee+2014,Lee2016,Lee+Park2018}. We attempted 
mode identification for the four independent frequencies detected in AX Dra with this method. The FRM is typically applicable to 
$\gamma$ Dor stars exhibiting at least three independent oscillation frequencies and having rotational velocities less than 70 km s$^{-1}$. 
Similar to the pulsating component of V404 Lyr \citep{Lee+2014}, the AX Dra A primary is a relatively fast rotator with 
$v_1 \sin i = 120\pm21$ km s$^{-1}$, well beyond this criterion. In addition, the pulsating component is likely affected by 
binary interaction, including mass transfer, which may further limit the applicability of the standard FRM approach because this method 
was originally developed under the assumption of single pulsating stars. 

Adopting the procedure of \citet{Lee+2014}, we set the frequency ratio error to a rather large value of $\pm$0.02 and looked for 
the model frequency ratios ($f_i$/$f_1$)$_{\rm model}$ that best fit the observed values of ($f_i$/$f_1$)$_{\rm obs}$. We found 
two acceptable sets of mode identifications within the adopted uncertainties, indicating that the FRM solutions are not unique. 
Rapid rotation and binary interaction may limit the applicability of the FRM method and distort the observed pulsation spectra. 
These two solutions are summarized in Table~\ref{Tab7}. Solution 1 corresponds to $(n,\ell)=(29,3)$, $(45,3)$, $(42,3)$, and $(40,3)$, 
while Solution 2 gives $(29,3)$, $(18,1)$, $(17,1)$, and $(16,1)$. The Brunt-V\"ais\"al\"a frequency integrals derived from 
Solutions 1 and 2 are 659.5$\pm$2.4 $\mu$Hz and 659.9$\pm$2.1 $\mu$Hz, respectively. Both values agree well with the model value of 
$\cal J_{\rm theo}$ $\approx$ 660 $\mu$Hz corresponding to physical parameters of $\log$ $T_{\rm eff}$ = 3.86, [Fe/H] = 0.0, and 
1.6 M$_\odot$ in the $\cal J -$ $\log$ $T_{\rm eff}$ diagram \citep{Moya+2005}. 

AX Dra belongs to the eclipsing $\gamma$ Dor star group and is one of the close binaries exhibiting both the shortest orbital period and 
the largest fill-out factor among the known members of this class \citep{Ibanoglu+2018,Gaulme+Guzik2019,Li+2020,Southworth+Van2022,Cakirli+2025}. 
In such a near-contact EB, the initially more massive component is expected to fill its Roche lobe and transfer mass to its companion, 
which then accretes the material and becomes the present more massive $\gamma$ Dor pulsator. In this context, mass transfer and 
tidal interactions between the AX Dra components may have contributed to the excitation of the $\gamma$ Dor oscillations identified in 
this study. The orbital-to-pulsation frequency ratio $f_{\rm orb}$:$f_{\rm 1}$ is close to, but not exactly, a 5:6 commensurability. 
This may indicate tidal interaction or near-resonance between the orbital motion and the dominant oscillation mode. Our results indicate 
that this system provides a promising laboratory for investigating the interplay between binarity and stellar pulsation. 


\begin{acknowledgements} 
This work includes echelle spectra obtained with BOES and 2-minute cadence photometric data from the TESS mission. Constructive comments 
from the anonymous referee greatly improved the clarity and quality of this paper. We acknowledge the use of the SIMBAD database operated 
at CDS, Strasbourg, France, and support from KASI grant 2026-1-904-01. 
\end{acknowledgements}

\clearpage

\clearpage 
\begin{deluxetable}{lcc}
\tablewidth{0pt}
\tablecaption{TESS Eclipse Times of AX Dra \label{Tab1}}
\tablehead{
\colhead{BJD}    & \colhead{Error} & \colhead{Min}  
}
\startdata
2,458,711.51517  & $\pm$0.00005    & I             \\
2,458,711.79868  & $\pm$0.00007    & II            \\
2,458,712.08286  & $\pm$0.00002    & I             \\
2,458,712.36813  & $\pm$0.00014    & II            \\
2,458,712.65058  & $\pm$0.00005    & I             \\
2,458,712.93626  & $\pm$0.00015    & II            \\
2,458,713.21897  & $\pm$0.00003    & I             \\
2,458,713.50403  & $\pm$0.00009    & II            \\
2,458,713.78772  & $\pm$0.00004    & I             \\
2,458,714.07127  & $\pm$0.00006    & II            \\
\enddata
\tablecomments{This table is available in its entirety in machine-readable form. A portion is shown here for guidance regarding its form and content.}
\end{deluxetable}

\begin{deluxetable}{lccccc}
\tablewidth{0pt}
\tablecaption{Light levels of AX Dra at four different phases \label{Tab2}}
\tablehead{
\colhead{Sector}  & \colhead{Median Time}  & \colhead{Min I at 0.0$P$} & \colhead{Max I at 0.25$P$} & \colhead{Min II at 0.5$P$} & \colhead{Max II at 0.75$P$}  \\
                  & \colhead{(BJD)}        & \colhead{(mag)}           & \colhead{(mag)}            & \colhead{(mag)}            & \colhead{(mag)}              
}
\startdata
15                & 2,458,724.38449        & 11.575$\pm$0.006          & 10.617$\pm$0.009           & 10.927$\pm$0.005           & 10.621$\pm$0.009             \\
21                & 2,458,884.11004        & 11.582$\pm$0.007          & 10.618$\pm$0.010           & 10.920$\pm$0.005           & 10.622$\pm$0.010             \\
22                & 2,458,912.90849        & 11.586$\pm$0.006          & 10.617$\pm$0.007           & 10.920$\pm$0.004           & 10.623$\pm$0.007             \\
41                & 2,459,433.28136        & 11.558$\pm$0.007          & 10.625$\pm$0.010           & 10.915$\pm$0.005           & 10.616$\pm$0.009             \\ 
                  & Average                & 11.575$\pm$0.011          & 10.619$\pm$0.003           & 10.921$\pm$0.004           & 10.620$\pm$0.003             \\
\enddata
\end{deluxetable}

\begin{deluxetable}{lrrrr}                                                                                            
\tablewidth{0pt}                    
\tabletypesize{\scriptsize}   
\tablecaption{Radial velocities of AX Dra$\dag$ \label{Tab3}}                                                                            
\tablehead{ 
\colhead{BJD}        & \colhead{$V_{1}$}       & \colhead{$\Delta V_{1}$} & \colhead{$V_{2}$}       & \colhead{$\Delta V_{2}$}  \\ 
                     & \colhead{(km s$^{-1}$)} & \colhead{(km s$^{-1}$)}  & \colhead{(km s$^{-1}$)} & \colhead{(km s$^{-1}$)}   
}
\startdata                                                                                                                       
2,459,732.0262       & $  -89.9  $             &  3.0                     & $ 190.5   $             &   8.2                     \\
2,459,732.0377       & $  -97.2  $             &  2.9                     & $ 205.5   $             &   8.9                     \\
2,459,732.0491       & $  -102.7 $             &  2.4                     & $ 199.1   $             &   12.1                    \\
2,459,732.0606       & $  -106.0 $             &  8.5                     & $ 210.7   $             &   22.9                    \\
2,459,732.0713       & $  -111.2 $             &  7.4                     & $ 214.7   $             &   14.0                    \\
2,459,732.0851       & $  -110.1 $             &  3.3                     & $ 224.8   $             &   11.7                    \\
2,459,732.0965       & $  -106.1 $             &  3.2                     & $ 222.7   $             &   2.4                     \\
2,459,732.1080       & $  -109.5 $             &  7.7                     & $ 208.7   $             &   21.8                    \\
2,459,732.1195       & $  -97.9  $             &  4.4                     & $ 191.0   $             &   15.1                    \\
2,459,732.1309       & $  -90.7  $             &  3.0                     & $ 187.5   $             &   5.1                     \\
2,459,732.1424       & $  -82.7  $             &  3.8                     & $ 182.6   $             &   11.0                    \\
2,459,733.1349       & $  -65.1  $             &  4.4                     & $ 148.0   $             &   8.8                     \\
2,459,733.1464       & $  -79.1  $             &  4.1                     & $ 161.7   $             &   9.9                     \\
2,459,734.0182       & $  100.4  $             &  4.2                     & $ -187.5  $             &   2.3                     \\
2,459,734.0898       & $  104.7  $             &  2.5                     & $ -219.7  $             &   2.9                     \\
2,459,734.1013       & $  101.8  $             &  1.8                     & $ -207.8  $             &   10.9                    \\
2,459,734.1242       & $  96.2   $             &  3.2                     & $ -185.1  $             &   4.5                     \\
2,459,734.1471       & $  86.3   $             &  1.5                     & $         $             &                           \\
2,459,741.9816       & $  99.3   $             &  4.5                     & $ -207.1  $             &   5.7                     \\
2,459,741.9931       & $  109.6  $             &  3.8                     & $ -214.4  $             &   5.8                     \\
2,459,742.0045       & $  104.7  $             &  3.7                     & $ -212.9  $             &   7.9                     \\
2,459,742.0160       & $  106.8  $             &  3.5                     & $ -224.6  $             &   4.8                     \\
2,459,742.0275       & $  111.9  $             &  2.5                     & $ -215.7  $             &   6.5                     \\
2,459,742.0618       & $  91.5   $             &  3.7                     & $ -208.8  $             &   8.0                     \\
2,459,742.0733       & $  90.0   $             &  4.0                     & $ -199.9  $             &   5.2                     \\
2,459,907.2752       & $  74.5   $             &  8.1                     & $ -146.6  $             &   9.5                     \\
2,459,907.2867       & $  86.7   $             &  5.5                     & $ -163.5  $             &   16.8                    \\
2,459,907.2982       & $  87.5   $             &  10.6                    & $ -185.8  $             &   15.1                    \\
2,459,907.3105       & $  104.1  $             &  2.5                     & $ -199.8  $             &   2.1                     \\
2,459,907.3220       & $  102.4  $             &  2.2                     & $ -214.3  $             &   3.1                     \\
2,459,907.3335       & $  102.0  $             &  2.3                     & $ -223.6  $             &   5.2                     \\
2,459,908.2753       & $  -85.9  $             &  3.5                     & $         $             &                           \\
2,459,908.2868       & $  -67.9  $             &  4.9                     & $ 164.6   $             &   8.2                     \\
2,459,908.2983       & $  -59.3  $             &  5.3                     & $ 150.7   $             &   10.2                    \\
2,459,908.3098       & $  -47.9  $             &  2.3                     & $ 128.0   $             &   8.8                     \\
2,459,908.3213       & $  -37.3  $             &  2.3                     & $         $             &                           \\
2,459,908.3327       & $  -24.3  $             &  2.2                     & $         $             &                           \\
2,459,908.3443       & $  -11.9  $             &  2.4                     & $         $             &                           \\
2,459,908.3558       & $  0.3    $             &  2.2                     & $         $             &                           \\
2,459,910.2835       & $  72.2   $             &  5.2                     & $ -155.2  $             &   6.8                     \\
2,459,910.2950       & $  72.4   $             &  5.6                     & $ -131.0  $             &   9.0                     \\
2,459,910.3065       & $  79.1   $             &  3.8                     & $ -92.5   $             &   9.6                     \\
2,459,910.3180       & $  77.6   $             &  4.2                     & $ -69.0   $             &   7.9                     \\
2,459,910.3295       & $  79.2   $             &  1.9                     & $ -49.2   $             &   3.7                     \\
2,459,910.3410       & $  30.2   $             &  1.7                     & $         $             &                           \\
2,459,910.3525       & $  -60.9  $             &  1.7                     & $         $             &                           \\
\enddata 
\tablenotetext{\dag}{$V_{1}$ and $V_{2}$ represent the measured RVs of the primary and secondary stars, respectively, and $\Delta V_{1}$ 
and $\Delta V_{2}$ are their 1$\sigma$-values. }
\end{deluxetable}

\begin{deluxetable}{lcc}
\tablewidth{0pt} 
\tabletypesize{\normalsize}   
\tablecaption{Light and Velocity Parameters of AX Dra \label{Tab4}}
\tablehead{
\colhead{Parameter}                & \colhead{Primary (A)}  & \colhead{Secondary (B)}      
} 
\startdata                                                                                                                            
$T_0$ (BJD)                        & \multicolumn{2}{c}{2,458,711.514718$\pm$0.000027}     \\
$P_{\rm orb}$ (day)                & \multicolumn{2}{c}{0.568163762$\pm$0.000000041}       \\
$a$ ($R_\odot$)                    & \multicolumn{2}{c}{3.927$\pm$0.025}                   \\
$\gamma$ (km s$^{-1}$)             & \multicolumn{2}{c}{$-$1.5$\pm$8.4}                    \\
$K_1$ (km s$^{-1}$)                & \multicolumn{2}{c}{111.6$\pm$1.3}                     \\
$K_2$ (km s$^{-1}$)                & \multicolumn{2}{c}{238.2$\pm$1.9}                     \\
$q$ (= $M_2/M_1$)                  & \multicolumn{2}{c}{0.4685$\pm$0.0066}                 \\
$i$ (deg)                          & \multicolumn{2}{c}{89.07$\pm$0.11}                    \\
$T_{\rm eff}$ (K)                  & 7220$\pm$150           & 4957$\pm$65                  \\
$\Omega$                           & 3.078$\pm$0.023        & 2.815                        \\
$F$                                & 0.87$\pm$0.15          & 1.0                          \\
$A$                                & 1.0                    & 0.573$\pm$0.033              \\
$g$                                & 1.0                    & 0.327$\pm$0.052              \\
$X_{\rm bol}$, $Y_{\rm bol}$       & 0.642, 0.259           & 0.643, 0.160                 \\
$x$, $y$                           & 0.505$\pm$0.010, 0.277 & 0.633$\pm$0.032, 0.184       \\
$l/(l_1+l_2)$                      & 0.8432$\pm$0.0057      & 0.1568                       \\
$r$ (pole)                         & 0.3804$\pm$0.0035      & 0.2947$\pm$0.0027            \\
$r$ (point)                        & 0.4233$\pm$0.0061      & 0.4227$\pm$0.0035            \\
$r$ (side)                         & 0.3917$\pm$0.0040      & 0.3075$\pm$0.0029            \\
$r$ (back)                         & 0.4065$\pm$0.0047      & 0.3400$\pm$0.0028            \\
$r$ (volume)$\rm ^a$               & 0.3934$\pm$0.0045      & 0.3153$\pm$0.0030            \\
\enddata
\tablenotetext{\rm ^a}{Mean volume radius. }
\end{deluxetable}

\begin{deluxetable}{lcc}
\tablewidth{0pt} 
\tablecaption{Absolute Parameters of AX Dra \label{Tab5}}
\tablehead{
\colhead{Parameter}                & \colhead{Primary (A)}  & \colhead{Secondary (B)}      
} 
\startdata                                                                                                                                 
$M$ ($M_\odot$)                    & 1.717$\pm$0.026        & 0.804$\pm$0.014             \\
$R$ ($R_\odot$)                    & 1.541$\pm$0.020        & 1.237$\pm$0.014             \\
$\log$ $g$ (cgs)                   & 4.297$\pm$0.013        & 4.158$\pm$0.013             \\
$\rho$ ($\rho_\odot$)              & 0.470$\pm$0.020        & 0.425$\pm$0.017             \\
$v_{\rm sync}$ (km s$^{-1}$)       & 137.3$\pm$1.8          & 110.2$\pm$1.3               \\
$v$$\sin$$i$ (km s$^{-1}$)         & 120$\pm$21             & \,                          \\
$T_{\rm eff}$ (K)                  & 7220$\pm$150           & 4957$\pm$65                 \\
$L$ ($L_\odot$)                    & 5.78$\pm$0.50          & 0.83$\pm$0.05               \\
$M_{\rm bol}$ (mag)                & 2.825$\pm$0.095        & 4.935$\pm$0.062             \\
BC (mag)                           & 0.034$\pm$0.001        & $-$0.312$\pm$0.027          \\
$M_{\rm V}$ (mag)                  & 2.791$\pm$0.095        & 5.247$\pm$0.068             \\
Distance (pc)                      & \multicolumn{2}{c}{381$\pm$19}                       \\
\enddata
\end{deluxetable}

\begin{deluxetable}{lrccrcc}
\tablewidth{0pt}
\tabletypesize{\small}   
\tablecaption{Multi-frequency analysis results for AX Dra$\rm ^a$ \label{Tab6}}
\tablehead{
                 & \colhead{Frequency}    & \colhead{Amplitude} & \colhead{Phase} & \colhead{S/N$\rm ^b$}  & \colhead{Remark}      \\
                 & \colhead{(day$^{-1}$)} & \colhead{(mmag)}    & \colhead{(rad)} &                        &
}                                                                                                      
\startdata                                                                                             
$f_{1}$$\rm ^c$  & 2.13680$\pm$0.00001    & 7.34$\pm$0.14       & 1.61$\pm$0.05   &   91.69                & $\gamma$ Dor$-$type   \\
$f_{2}$$\rm ^c$  & 1.38336$\pm$0.00001    & 3.93$\pm$0.15       & 1.28$\pm$0.11   &   45.92                & $\gamma$ Dor$-$type   \\
$f_{3}$$\rm ^c$  & 1.47095$\pm$0.00002    & 1.97$\pm$0.15       & 6.19$\pm$0.22   &   23.04                & $\gamma$ Dor$-$type   \\
$f_{4}$$\rm ^c$  & 0.75348$\pm$0.00003    & 1.37$\pm$0.15       & 1.84$\pm$0.32   &   15.49                & $f_1-f_2$             \\
$f_{5}$$\rm ^c$  & 1.55140$\pm$0.00003    & 1.22$\pm$0.15       & 2.60$\pm$0.35   &   14.42                & $\gamma$ Dor$-$type   \\
$f_{6}$          & 2.15100$\pm$0.00003    & 1.15$\pm$0.14       & 3.92$\pm$0.35   &   14.44                & $3f_3-3f_4$           \\
$f_{7}$          & 1.49282$\pm$0.00003    & 1.26$\pm$0.15       & 1.30$\pm$0.34   &   14.79                &                       \\
$f_{8}$          & 2.12282$\pm$0.00004    & 0.86$\pm$0.14       & 1.51$\pm$0.47   &   10.69                & $2f_1-f_6$            \\
$f_{9}$$\rm ^c$  & 3.14339$\pm$0.00003    & 1.00$\pm$0.12       & 4.67$\pm$0.35   &   14.50                & $f_{\rm orb}+f_2$     \\
$f_{10}$$\rm ^c$ & 4.27360$\pm$0.00003    & 0.74$\pm$0.09       & 4.23$\pm$0.37   &   13.62                & $2f_1$                \\
$f_{11}$         & 2.09611$\pm$0.00004    & 0.83$\pm$0.14       & 6.20$\pm$0.48   &   10.34                & $2f_8-f_6$            \\
$f_{12}$         & 1.39900$\pm$0.00005    & 0.81$\pm$0.15       & 5.55$\pm$0.53   &    9.45                & $f_6-f_4$             \\
$f_{13}$         & 2.28151$\pm$0.00005    & 0.71$\pm$0.13       & 6.05$\pm$0.55   &    9.04                &                       \\
$f_{14}$         & 2.17617$\pm$0.00005    & 0.78$\pm$0.14       & 1.02$\pm$0.52   &    9.73                & $f_{10}-f_{11}$       \\
$f_{15}$         & 2.06778$\pm$0.00007    & 0.53$\pm$0.14       & 1.06$\pm$0.77   &    6.54                &                       \\
$f_{16}$         & 2.22440$\pm$0.00007    & 0.51$\pm$0.14       & 4.08$\pm$0.77   &    6.49                & $f_3+f_4$             \\
$f_{17}$         & 0.66580$\pm$0.00008    & 0.49$\pm$0.15       & 4.29$\pm$0.91   &    5.54                & $f_1-f_3$             \\ 
$f_{18}$         & 1.99221$\pm$0.00009    & 0.43$\pm$0.14       & 2.06$\pm$0.93   &    5.38                & $f_{10}-f_{13}$       \\ 
$f_{19}$         & 2.34501$\pm$0.00007    & 0.55$\pm$0.13       & 5.24$\pm$0.70   &    7.11                &                       \\
$f_{20}$         & 1.60747$\pm$0.00008    & 0.48$\pm$0.14       & 5.55$\pm$0.88   &    5.72                &                       \\
$f_{21}$         & 2.04387$\pm$0.00008    & 0.48$\pm$0.14       & 1.47$\pm$0.85   &    5.89                &                       \\
$f_{22}$         & 1.00656$\pm$0.00008    & 0.52$\pm$0.15       & 5.43$\pm$0.83   &    6.05                & $f_9-f_1$             \\
$f_{23}$         & 1.67971$\pm$0.00010    & 0.38$\pm$0.14       & 1.82$\pm$1.11   &    4.52                & $f_{19}-f_{17}$       \\
$f_{24}$         & 1.53669$\pm$0.00007    & 0.53$\pm$0.15       & 0.61$\pm$0.80   &    6.23                & $f_9-f_{20}$          \\
$f_{25}$         & 1.37443$\pm$0.00008    & 0.50$\pm$0.15       & 3.16$\pm$0.86   &    5.83                &                       \\
$f_{26}$         & 2.89028$\pm$0.00009    & 0.39$\pm$0.12       & 5.48$\pm$0.93   &    5.41                & $f_1+f_4$             \\
$f_{27}$         & 1.96910$\pm$0.00009    & 0.43$\pm$0.14       & 0.45$\pm$0.95   &    5.28                &                       \\
$f_{28}$         & 0.76773$\pm$0.00010    & 0.42$\pm$0.15       & 1.43$\pm$1.06   &    4.74                & $f_6-f_2$             \\
$f_{29}$         & 0.30798$\pm$0.00009    & 0.44$\pm$0.15       & 3.10$\pm$1.02   &    4.93                & $f_{15}-f_{\rm orb}$  \\
$f_{30}$         & 2.20688$\pm$0.00007    & 0.55$\pm$0.14       & 2.72$\pm$0.73   &    6.87                & $f_{10}-f_{15}$       \\
$f_{31}$         & 0.07357$\pm$0.00011    & 0.37$\pm$0.15       & 4.46$\pm$1.21   &    4.13                & $f_3-f_{12}$          \\
$f_{32}$         & 2.53294$\pm$0.00010    & 0.36$\pm$0.13       & 3.73$\pm$1.05   &    4.76                & $f_{16}+f_{29}$       \\
$f_{33}$         & 1.78101$\pm$0.00011    & 0.34$\pm$0.14       & 5.59$\pm$1.22   &    4.12                & $f_{32}-f_4$          \\
$f_{34}$         & 1.50777$\pm$0.00009    & 0.43$\pm$0.15       & 5.62$\pm$0.98   &    5.11                & $2f_4$                \\
$f_{35}$         & 2.23690$\pm$0.00010    & 0.37$\pm$0.14       & 5.04$\pm$1.06   &    4.71                & $f_{28}+f_3$          \\
\enddata                                                                                                                           
\tablenotetext{\rm ^a}{Frequencies are listed in order of detection. }
\tablenotetext{\rm ^b}{Calculated in a range of 5 day$^{-1}$ around each frequency. }
\tablenotetext{\rm ^c}{Detected in all four sectors. }
\end{deluxetable}

\newpage
\begin{deluxetable}{lccccccc}
\tabletypesize{\scriptsize} 
\tablewidth{0pt}
\tablecaption{$\gamma$ Dor-type Pulsation Properties of AX Dra \label{Tab7}}
\tablehead{
   & \colhead{Frequency}    & \colhead{$Q$}    & \colhead{($f_i$/$f_1$)$_{\rm obs}$} & \colhead{mode ($n$, $\ell$)}  & \colhead{($f_i$/$f_1$)$_{\rm model}$} & \colhead{$\Delta$($f_i$/$f_1$)$_{\rm obs-model}$} & \colhead{$\cal J_{\rm obs}$}    \\ [-1.0ex]
   & \colhead{(day$^{-1}$)} & \colhead{(days)} &                                     &                               &                                       &                                                   & \colhead{($\mu$Hz)}  
}
\startdata		
\multicolumn{8}{l}{FRM Solution 1} \\
$f_1$    & 2.13680   & 0.320  & \,       & (29, 3)    & \,       & \,                     & 661.65            \\	
$f_2$    & 1.38336   & 0.494  & 0.6474   & (45, 3)    & 0.6484   & $-$0.0010              & 660.68            \\		
$f_3$    & 1.47095   & 0.465  & 0.6884   & (42, 3)    & 0.6941   & $-$0.0057              & 656.19            \\	
$f_5$    & 1.55140   & 0.440  & 0.7260   & (40, 3)    & 0.7284   & $-$0.0024              & 659.51            \\
         &           &        &          &            & Average  & $-$0.0030$\pm$0.0025   & 659.51$\pm$2.38   \\ \tableline
\multicolumn{8}{l}{FRM Solution 2} \\
$f_1$    & 2.13680   & 0.320  & \,       & (29, 3)    & \,       & \,                     & 661.65            \\	
$f_2$    & 1.38336   & 0.494  & 0.6474   & (18, 1)    & 0.6510   & $-$0.0036              & 658.00            \\		
$f_3$    & 1.47095   & 0.465  & 0.6884   & (17, 1)    & 0.6882   & $-$0.0002              & 661.84            \\	
$f_5$    & 1.55140   & 0.440  & 0.7260   & (16, 1)    & 0.7299   & $-$0.0039              & 658.15            \\
         &           &        &          &            & Average  & $-$0.0024$\pm$0.0023   & 659.91$\pm$2.12   \\         
\enddata                                                                                                                           
\end{deluxetable}


\newpage
\begin{thebibliography}{}
\providecommand{\dodoi}[1]{doi:~\href{http://doi.org/#1}{\nolinkurl{#1}}}
\providecommand{\doarXiv}[1]{\href{https://arxiv.org/abs/#1}{\nolinkurl{https://arxiv.org/abs/#1}}}

\bibitem[V. Antoci et al.(2019)]{Antoci+2019} Antoci, V., Cunha, M. S., Bowman, D. M., et al. 2019, MNRAS, 490, 4040, \dodoi{10.1093/mnras/stz2787}
\bibitem[S. Blanco-Cuaresma(2019)]{Blanco-cuaresma2019} Blanco-Cuaresma, S. 2019, MNRAS, 486, 2075, \dodoi{10.1093/mnras/stz549}
\bibitem[S. Blanco-Cuaresma et al.(2014)]{Blanco-Cuaresma+2014} Blanco-Cuaresma, S., Soubiran, C., Heiter, U., et al. 2014, A\&A, 569, A111, \dodoi{10.1051/0004-6361/201423945}
\bibitem[M. Breger(2000)]{Breger2000} Breger, M. 2000, in ASP Conf. Ser. 210, Delta Scuti and Related Stars, ed. M. Breger, \& M. H. Montgomery (San Francisco, CA: ASP), 3
\bibitem[M. Breger \& K. M. Bischof(2002)]{Breger+Bischof2002} Breger, M., \& Bischof, K. M. 2002, A\&A, 385, 537, \dodoi{10.1051/0004-6361:20020124}
\bibitem[M. Breger et al.(1993)]{Breger+1993} Breger, M., Stich, J., Garrido, R., et al. 1993, A\&A, 271, 482, \dodoi{10.1051/0004-6361:1993271482}
\bibitem[\" O. \c Cakirli et al.(2025)]{Cakirli+2025} \c Cakirli, \" O., Hoyman, B., O\"zdarcan, O., \& Bilir, S. 2025, MNRAS, 533, 2058 \dodoi{10.1093/mnras/staf330} 
\bibitem[P. Coelho et al.(2005)]{Coelho+2005} Coelho, P., Barbuy, B., Melendez, J., Sciavon, R. P., \& Castilho, B. V. 2005, A\&A, 443, 735, \dodoi{10.1051/0004-6361:20053511} 
\bibitem[P. J. Flower(1996)]{Flower1996} Flower, P. J. 1996, ApJ, 469, 355, \dodoi{10.1086/177785}
\bibitem[D. P. Fleming et al.(2019)]{Fleming+2019} Fleming, D. P., Barnes, R., Davenport, J. R. A., \& Luger, R. 2019, ApJ, 881, 88, \dodoi{10.3847/1538-4357/ab2ed2}
\bibitem[J. Fuller \& C. Felce(2023)]{Fuller+Felce2023} Fuller, J., \& Felce, C. 2023, MNRAS, 527, L103, \dodoi{10.1093/mnras/stad3053}
\bibitem[Gaia Collaboration et al.(2023)]{Gaia2023} Gaia Collaboration, Vallenari A., Brown A. G. A., et al. 2023, A\&A, 674, A1, \dodoi{10.1051/0004-6361/202243940} 
\bibitem[P. Gaulme \& J. A. Guzik(2019)]{Gaulme+Guzik2019} Gaulme, P., \& Guzik, J. A. 2019, A\&A, 630, A106, \dodoi{10.1051/0004-6361/201935821} 
\bibitem[R. O. Gray \& C. J. Corbally(1994)]{Gray+Corbally1994} Gray, R. O., \& Corbally, C. J. 1994, AJ, 107, 742, \dodoi{10.1086/116893}
\bibitem[R. O. Gray \& C. J. Corbally(2009)]{Gray+Corbally2009} Gray, R. O., \& Corbally, C. J. 2009, Stellar Spectral Classification (Princeton, NJ: Princeton Univ. Press)
\bibitem[A. Grigahc\'ene et al.(2010)]{Grigahcene+2010} Grigahc\'ene, A., Antoci, V., Balona, L., et al. 2010, ApJ, 713, L192, \dodoi{10.1088/2041-8205/713/2/L192}
\bibitem[G. Handler \& R. R. Shobbrook(2002)]{Handler+Shobbrook2002} Handler, G., \& Shobbrook, R. R. 2002, MNRAS, 333, 251, \dodoi{10.1046/j.1365-8711.2002.05401.x}
\bibitem[U. Heiter et al.(2015)]{Heiter+2015} Heiter, U., Lind, K., Asplund, M., et al. 2015, PhyS, 90, 054010, \dodoi{10.1088/0031-8949/90/5/05401}
\bibitem[G. W. Henry et al.(2005)]{Henry+2005} Henry, G. W., Fekel, F. C., \& Henry, S. M. 2005, AJ, 129, 2815, \dodoi{10.1086/429876}
\bibitem[M. Hobson-Ritz et al.(2025)]{Hobson-Ritz+2025} Hobson-Ritz, M., Birky, J., Peterson, L., et al. 2025, ApJ, 990, 124, \dodoi{10.3847/1538-4357/adf10d} 
\bibitem[E. H\o g et al.(2000)]{Hog+2000} H\o g, E., Fabricius, C., Makarov, V. V., et al. 2000, A\&A, 355, 27, \dodoi{10.1051/0004-6361:20020249} 
\bibitem[K. Hong et al.(2015)]{Hong+2015} Hong, K., Lee, J. W., Kim, S.-L., et al. 2015, AJ, 150, 131, \dodoi{10.1088/0004-6256/150/4/131}
\bibitem[S. Iliji\'c et al.(2004)]{Ilijic+2004} Iliji\'c, S., Hensberge, H., Pavlovski, K., \& Freyhammer, L. M. 2004, in ASP Conf. Ser. 318, Spectroscopically and Spatially Resolving the Components of the Close Binary Stars, ed. R. Hilditch, H. Hensberge, \& K. Pavlovski (San Francisco: ASP), 111
\bibitem[C. Ibanoglu et al.(2018)]{Ibanoglu+2018} Ibanoglu, C., \c Cakirli, \" O., \& Sipahi, E., 2018, New Astron., 62, 70, \dodoi{10.1016/j.newast.2018.01.004}
\bibitem[H. I. Kim et al.(2004)]{Kim+2004} Kim, H. I., Lee, J. W., Kim, C. H., et al. 2004, PASP, 116, 931, \dodoi{10.1086/425606}
\bibitem[K.-M. Kim et al.(2007)]{Kim+2007} Kim, K.-M., Han, I., Valyavin, G. G., et al. 2007, PASP, 119, 1052, \dodoi{10.1086/521959} 
\bibitem[R. Kippenhahn(1955)]{Kippenhahn1955} Kippenhahn, R. 1955, Kl. Ver\"off. Remeis-Sternwarte, Bamberg, No. 9
\bibitem[G. Koenigsberger et al.(2021)]{Koenigsberger+2021} Koenigsberger, G., Moreno, E., \& Langer, N. 2021, A\&A, 653, A127, \dodoi{10.1051/0004-6361/202039369}
\bibitem[D. W. Kurtz(2022)]{Kurtz2022} Kurtz, D. W. 2022, ARAA, 60, 31, \dodoi{10.1146/annurev-astro-052920-094232}
\bibitem[R. L. Kurucz(2005)]{kurucz2005} Kurucz, R. L. 2005, MSAIS, 8, 14
\bibitem[K. K. Kwee \& H. Van Woerden(1956)]{Kwee+Van1956} Kwee, K. K., \& Van Woerden, H. 1956, Bull. Astron. Inst. Netherlands, 12, 327
\bibitem[J. W. Lee(2016)]{Lee2016} Lee, J. W. 2016, ApJ, 833, 170, \dodoi{10.3847/1538-4357/833/2/170}
\bibitem[J. W. Lee \& J.-H. Park(2018)]{Lee+Park2018} Lee, J. W., \& Park, J.-H. 2018, MNRAS, 480, 4693, \dodoi{10.1093/mnras/sty2153} 
\bibitem[J. W. Lee et al.(2021)]{Lee+2021} Lee, J. W., Hong, K., \& Kim, H.-Y. 2021, AJ, 161, 129, \dodoi{10.3847/1538-3881/abd631}
\bibitem[J. W. Lee et al.(2022)]{Lee+2022} Lee, J. W., Hong, K., Kim, H.-Y., \& Park, J.-H. 2022, MNRAS, 515, 4702, \dodoi{10.1093/mnras/stac2151} 
\bibitem[J. W. Lee et al.(2018)]{Lee+2018} Lee, J. W., Hong, K., Koo, J.-R., \& Park J.-H., 2018, AJ, 155, 5, \dodoi{10.3847/1538-3881/aa947e} 
\bibitem[J. W. Lee et al.(2023)]{Lee+2023} Lee, J. W., Hong, K., Park, J.-H., Wolf, M., \& Kim, D.-J. 2023, AJ, 165, 159, \dodoi{10.3847/1538-3881/acbe9d}
\bibitem[J. W. Lee et al.(2014)]{Lee+2014} Lee, J. W., Kim, S.-L., Hong, K., Lee, C.-U., \& Koo, J.-R. 2014, AJ, 148, 37, \dodoi{10.1088/0004-6256/148/2/37} 
\bibitem[J. W. Lee et al.(2019)]{Lee+2019} Lee, J. W., Kristiansen, M., \& Hong, K. 2019, AJ, 157, 223, \dodoi{10.3847/1538-3881/ab1a3b} 
\bibitem[P. Lenz \& M. Breger(2005)]{Lenz+Breger2005} Lenz, P., \& Breger, M. 2005, Comm. Asteroseismology, 146, 53, \dodoi{10.1553/cia146s53} 
\bibitem[G. Li et al.(2020)]{Li+2020} Li, G., Guo, Z., Fuller, J., et al. 2020, MNRAS, 497, 4363, \dodoi{10.1093/mnras/staa2266}
\bibitem[G. L. Loumos \&  T. J. Deeming(1978)]{Loumos+Deeming1978} Loumos, G. L., \& Deeming, T. J. 1978, Ap\&SS, 56, 285, \dodoi{10.1007/BF01879560}
\bibitem[J. C. Lurie et al.(2017)]{Lurie+2017} Lurie, J. C., Vyhmeister, K., Hawley, S. L., et al. 2017, AJ, 154, 250, \dodoi{10.3847/1538-3881/aa974d}
\bibitem[A. Moya et al.(2005)]{Moya+2005} Moya, A., Su\'arez, J. C., Amado P. J., et al. 2005, A\&A, 432, 189, \dodoi{10.1051/0004-6361:20041752}
\bibitem[M. Paegert et al.(2022)]{Paegert+2022} Paegert, M., Stassun, K. G., Collins, K. A., et al. 2022, VizieR Online Data Catalog, IV/39 
\bibitem[M. J. Pecaut \& E. E. Mamajek(2013)]{Pecaut+Mamajek2013} Pecaut, M. J., \& Mamajek, E. E. 2013, ApJS, 208, 9, \dodoi{10.1088/0067-0049/208/1/9}
\bibitem[B. Pilecki et al.(2017)]{Pilecki+2017} Pilecki, B., Gieren, W., Smolec, S., et al. 2017, ApJ, 842, 110, \dodoi{10.3847/1538-4357/aa6ff7}
\bibitem[P. Rittipruk et al.(2025)]{Rittipruk+2025} Rittipruk, P., Hong, K., Lee, J. W., et al. 2025, AJ, 169, 66, \dodoi{10.3847/1538-3881/ad99ce} 
\bibitem[S. Rucinski(1999)]{Rucinski1999} Rucinski, S. 1999, in ASP Conf. Ser. 185, Precise Stellar Radial Velocities, ed. J. B. Hearnshaw \& C. D. Scarfe (San Francisco, CA: ASP), 82, \dodoi{10.48550/arXiv.astro-ph/9807327} 
\bibitem[S. M. Rucinski(1992)]{Rucinski1992} Rucinski, S. M. 1992, AJ, 104, 1968, \dodoi{10.1086/181071}
\bibitem[S. M. Rucinski(2002)]{Rucinski2002} Rucinski, S. M. 2002, AJ, 124, 1746, \dodoi{10.1086/342342} 
\bibitem[J. F. Sepinsky et al.(2007)]{Sepinsky+2007} Sepinsky, J. F., Willems, B., Kalogera, V., \& Rasio, F. A. 2007, ApJ, 667, 1170, \dodoi{10.1086/520911}
\bibitem[J. F. Sepinsky et al.(2010)]{Sepinsky+2010} Sepinsky, J. F., Willems, B., Kalogera, V., \& Rasio, F. A. 2010, ApJ, 724, 546, \dodoi{10.1088/0004-637X/724/1/546}
\bibitem[J. Southworth \& D. M. Bowman(2022)]{Southworth+Bowman2022} Southworth, J., \& Bowman, D. M. 2022, The Observatory, 142, 161, \dodoi{10.48550/arXiv.2205.08841}
\bibitem[J. Southworth \& T. Van Reeth(2022)]{Southworth+Van2022} Southworth, J., \& Van Reeth, T., 2022, MNRAS, 515, 2755, \dodoi{10.1093/mnras/stac1993}
\bibitem[W. Strohmeier \& R. Knigge(1961)]{Strohmeier+Knigge1961} Strohmeier, W., \& Knigge, R. 1961, Ver\"off. Remeis-Sternwarte, Bamberg 5, No. 11, 3
\bibitem[T. M. Tauris \& E. P. J. van den Heuvel(2006)]{Tauris+van2006} Tauris, T. M., \& van den Heuvel, E. P. J. 2006, in in Compact stellar X-ray
sources, ed. W. H. G. Lewin \& M. van der Klis (Cambridge: Cambridge Univ. Press), 623 \dodoi{10.48550/arXiv.astro-ph/0303456}
\bibitem[G. Torres et al.(2010)]{Torres+2010} Torres, G., Andersen, J., \& Gim\'enez, A. 2010, A\&AR, 18, 67, \dodoi{10.1007/s00159-009-0025-1}
\bibitem[K. Uytterhoeven et al.(2011)]{Uytterhoeven+2011} Uytterhoeven, K., Moya, A., Grigahc\'ene, A., et al. 2011, A\&A, 534, A125, \dodoi{10.1051/0004-6361/201117368}
\bibitem[W. Van Hamme(1993)]{Van1993} Van Hamme, W. 1993, AJ, 106, 209, \dodoi{10.1086/116788} 
\bibitem[W. Van Hamme \& R. E. Wilson(2007)]{Van+Wilson2007} Van Hamme, W., \& Wilson, R. E. 2007, ApJ, 661, 1129, \dodoi{10.1086/517870}
\bibitem[R. E. Wilson \& E. J. Devinney(1971)]{Wilson+Devinney1971} Wilson, R. E., \& Devinney, E. J. 1971, ApJ, 166, 605, \dodoi{10.1086/150986} 
\bibitem[D. R. Xiong et al.(2016)]{Xiong+2016} Xiong, D. R., Deng, L., Zhang, C., et al. 2016, MNRAS, 457, 3163, \dodoi{10.1093/mnras/stw047} 
\bibitem[X. B. Zhang et al.(2013)]{Zhang+2013} Zhang, X. B., Luo, C. Q., \& Fu, J. N. 2013, ApJ, 777, 77, \dodoi{10.1088/0004-637X/777/1/77} 
\end{thebibliography}
\end{document}